\documentclass[12pt]{article}
\usepackage[style=chem-acs,articletitle=true]{biblatex}
\usepackage[parfill]{parskip}

\usepackage{geometry}
\usepackage{float}
\usepackage{siunitx}
\usepackage{upgreek}
\usepackage{setspace}
\usepackage{etoc}
\newcommand{\bfB}{\mathbf{B}}
\newcommand{\bfk}{\mathbf{k}}
\newcommand{\bfm}{\mathbf{m}}
\newcommand{\ux}{\mathbf{\hat{x}}}
\newcommand{\uy}{\mathbf{\hat{y}}}
\newcommand{\uz}{\mathbf{\hat{z}}}
 \usepackage[font=small,font=sf, labelfont=bf]{caption}

\newcommand{\beginsupplement}{%
        \setcounter{section}{0}
        \renewcommand{\thesubsection}{S\arabic{subsection}.}%
        \setcounter{table}{0}
        \renewcommand{\thetable}{S\arabic{table}}%
        \setcounter{figure}{0}
        \renewcommand{\thefigure}{S\arabic{figure}}%
        \setcounter{equation}{0}
        \renewcommand{\theequation}{S\arabic{equation}}%
     }

\usepackage[utf8]{inputenc}
\usepackage{mathtools}
\usepackage{amsmath}
\usepackage{braket}
\usepackage{textcomp}
\usepackage{xcolor}
\usepackage{hyperref}
\usepackage{bm}
\def\be{\begin{equation}}
\def\ee{\end{equation}}
\title{\Large{\textbf{Broadband microwave detection using electron spins in a hybrid diamond-magnet sensor chip}}}
\author{{\normalsize Joris J. Carmiggelt}$^{1}$, {\normalsize Iacopo Bertelli}$^{1}$, {\normalsize Roland W. Mulder}$^{1}$, {\normalsize Annick Teepe}$^{1}$, {\normalsize Mehrdad Elyasi}$^{2}$, \\ {\normalsize Brecht G. Simon}$^{1}$, {\normalsize Gerrit E. W. Bauer}$^{1,2,3}$, {\normalsize Yaroslav M. Blanter}$^{1}$, {\normalsize Toeno van der Sar}$^{1,*}$}
\date{
\footnotesize{
\raggedright \parindent=15pt $^1$Department of Quantum Nanoscience, Kavli Institute of Nanoscience, Delft University of Technology; \\ \raggedright \parindent=15pt Lorentzweg 1, 2628 CJ Delft, The Netherlands. \\
$^2$Advanced institute for Materials Research (WPI-AIMR), Tohoku University; Sendai 980-8577, Japan.\\
$^3$Kavli Institute for Theoretical Sciences, University of Chinese Academy of Sciences; Beijing 100190, China. \\
\hfill\break
$^*$Corresponding author. Email: \href{mailto:T.vanderSar@tudelft.nl}{T.vanderSar@tudelft.nl} }
}
\begin{document}
\maketitle
\begin{abstract}
\noindent Quantum sensing has developed into a main branch of quantum science and technology. It aims at measuring physical quantities with high resolution, sensitivity, and dynamic range. Electron spins in diamond are powerful magnetic field sensors, but their sensitivity in the microwave regime is limited to a narrow band around their resonance frequency. Here, we realize broadband microwave detection using spins in diamond interfaced with a thin-film magnet. A pump field locally converts target microwave signals to the sensor-spin frequency via the non-linear spin-wave dynamics of the magnet. Two complementary conversion protocols enable sensing and high-fidelity spin control over a gigahertz bandwidth, allowing characterization of the spin-wave band at multiple gigahertz above the sensor-spin frequency. The pump-tunable, hybrid diamond-magnet sensor chip opens the way for spin-based sensing in the 100-gigahertz regime at small magnetic bias fields. 
\end{abstract}
\newpage
\begin{refsection}
\addcontentsline{toc}{section}{Main Text}Electron spins associated with nitrogen-vacancy (NV) defects in diamond are magnetic field sensors that provide high spatial resolution and sensitivity at room temperature~\cite{Degen2017, Rondin2014}. They have been used to study nuclear magnetic resonance at the nanoscale~\cite{Aslam2017, Lovchinsky2017}, bio-~\cite{Schirhagl2014}, paleo-~\cite{Glenn2017}, and solid-state magnetism~\cite{Casola2018}, and electric currents in quantum materials~\cite{Ku2020, Tetienne2017}. Most of these applications focus on detecting magnetic fields in the 0-100 megahertz (MHz) frequency range, in which a toolbox of spin-control techniques enables high sensitivity and a tunable detection frequency without requiring a specific electron spin resonance (ESR) frequency~\cite{Degen2017}. In contrast, NV-based sensing in the microwave regime [1-100 gigahertz (GHz)] currently relies on tuning the ESR to the frequency of interest using a magnetic bias field~\cite{Appel2015}. This bias field changes the properties of e.g. magnetic or superconducting samples under study~\cite{Bertelli2020, Thiel2016}, for instance by altering their excitation spectrum, which limits its application in materials science. Furthermore, the field must be on the Tesla scale for operation in the 10-100 GHz range~\cite{Fortman2021}, making the required magnets large and slow to adjust, precluding the small sensor packaging desired for technological applications. \\
Here, we enable broadband spin-based microwave sensing by interfacing a diamond chip containing a layer of NV sensor spins with a thin-film magnet. The central concept is that the non-linear dynamics of spin waves - the collective spin excitations of the magnetic film~\cite{Chumak2015} - locally convert a target signal to the NV ESR frequency under the application of a pump field (Fig. 1A-B). We realize a $\sim$1-GHz detection bandwidth at fixed magnetic bias field via four-spin-wave mixing, and microwave detection at multiple GHz above the ESR frequency via difference-frequency generation. The pump-tunable detection frequency enables characterizing the spin-wave band structure despite a multi-GHz detuning and provides insight into the non-linear spin-wave dynamics limiting the conversion process. Furthermore, the converted microwaves are highly coherent, enabling high-fidelity control of the sensor spins via off-resonant drive fields. \\
Our hybrid diamond-magnet sensor platform consists of an ensemble of near-surface NV spins in a diamond membrane positioned onto a thin film of yttrium iron garnet (YIG) - a magnetic insulator with low spin-wave damping~\cite{Chumak2015} (Fig. 1B). A stripline delivers the “two-color” signal and pump microwave fields to the YIG film, in which they excite spin waves at the signal and pump frequencies, $f_\text{s}$ and $f_\text{p}$, respectively. The frequency-converted microwaves at the ESR frequency $f_\text{NV}$ are detected by measuring the spin-dependent NV photoluminescence under green laser excitation (Methods and Fig. 1C). The ESR frequency is fixed by an external magnetic bias field $B_\text{NV}$ (Fig. 1D). \\
\begin{figure}[h!]\centering
	\includegraphics[scale=1.3]{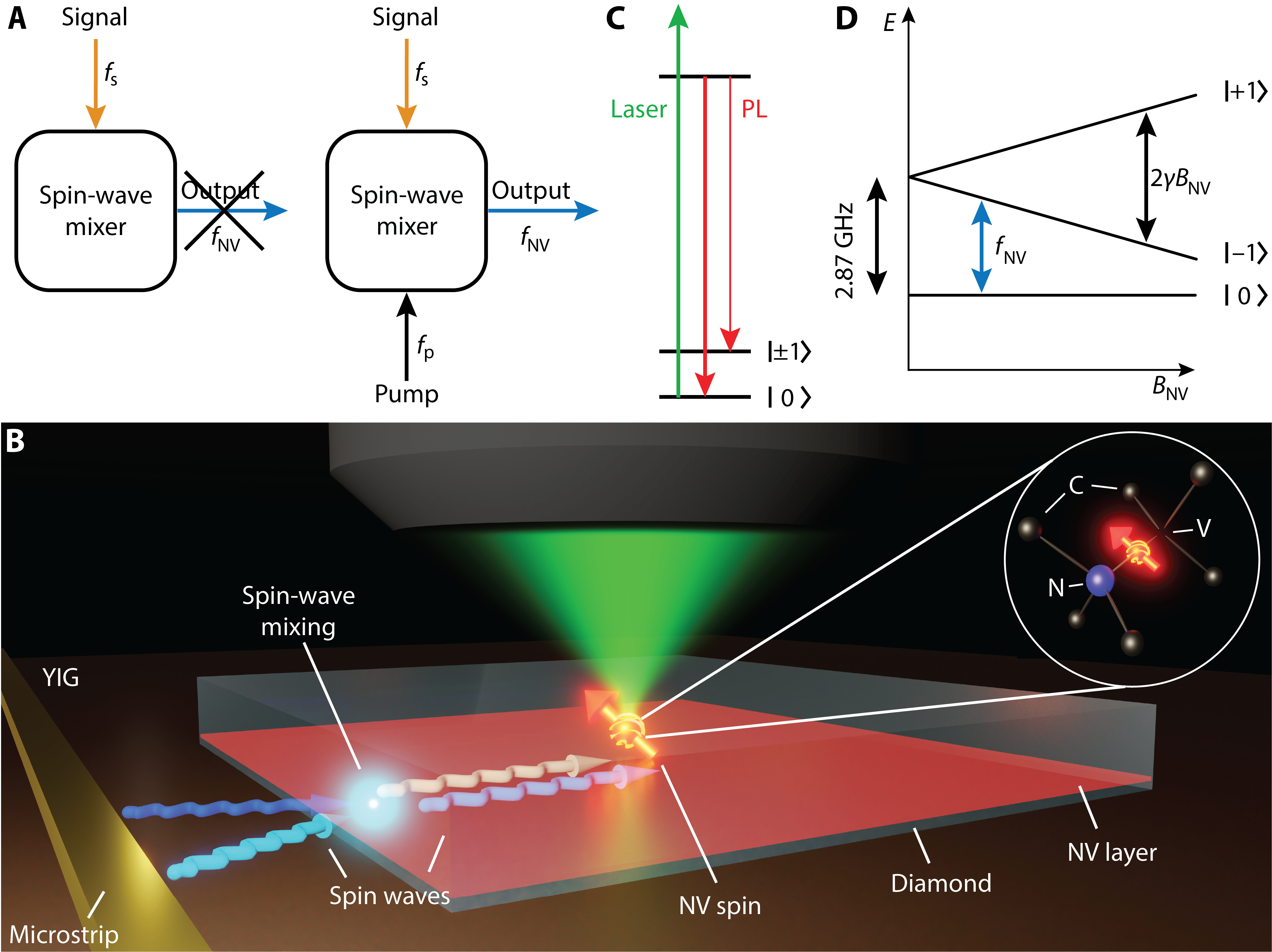}
	\caption{\textbf{Detecting microwave magnetic fields using spins in diamond via on-chip spin-wave-mediated frequency conversion.} (\textbf{A}) Idea of the experiment. A “spin-wave mixer” uses a pump to convert a microwave signal at frequency $f_\text{s}$ to an output frequency $f_\text{NV}$ that is detectable by nitrogen-vacancy (NV) sensor spins in diamond. (\textbf{B}) Sketch of the setup. A diamond with NV centers implanted $\sim$ 10-20 nm below its surface is placed onto a film of yttrium iron garnet (YIG, thickness: 235 nm). A microstrip delivers the signal and pump microwaves, which excite spin waves in the YIG. Spin-wave mixing enables detection of the signal field by converting its frequency to the NV electron spin resonance (ESR) frequency. Inset: Atomic structure of an NV center in the diamond carbon (C) lattice. (\textbf{C}) Initialization and readout of the NV spins is achieved through excitation by a green laser and detection of the red photoluminescence (PL). The PL is stronger in the $m_s=\ket{0}$ state than in the $m_s=\ket{\pm1}$ states. (\textbf{D}) NV spin energy levels in the electronic ground state. A magnetic field $B_\text{NV}$ along the NV axis splits the $m_s=\ket{\pm1}$ states via the Zeeman interaction. From the four possible configurations in the diamond lattice, we use the NV orientation with in-plane projection parallel to the stripline. $f_\text{NV}$ denotes the $\ket{0}\leftrightarrow\ket{-1}$ ESR transition frequency.                
	}
	\label{fig1}
\end{figure}
Our first detection protocol harnesses degenerate four-spin-wave mixing~\cite{Marsh2012,Suhl1957,Schultheiss2012} - the magnetic analogue of optical four-wave mixing (Fig. 2A). In the quasiparticle picture, this process corresponds to the scattering of two “pump” magnons into a “signal” magnon and an “idler” magnon at frequency $f_\text{i}=2f_\text{p}-f_\text{s}$. This conversion enables the detection of a microwave signal that is detuned from the ESR frequency, which would be otherwise invisible in the optical response of the NV centers (Fig. 2B). By tuning the frequency of the pump, we enable the detection of signals of specific microwave frequencies (Fig. 2C). \\
\begin{figure}[h!]\centering
	\includegraphics[scale=1.3]{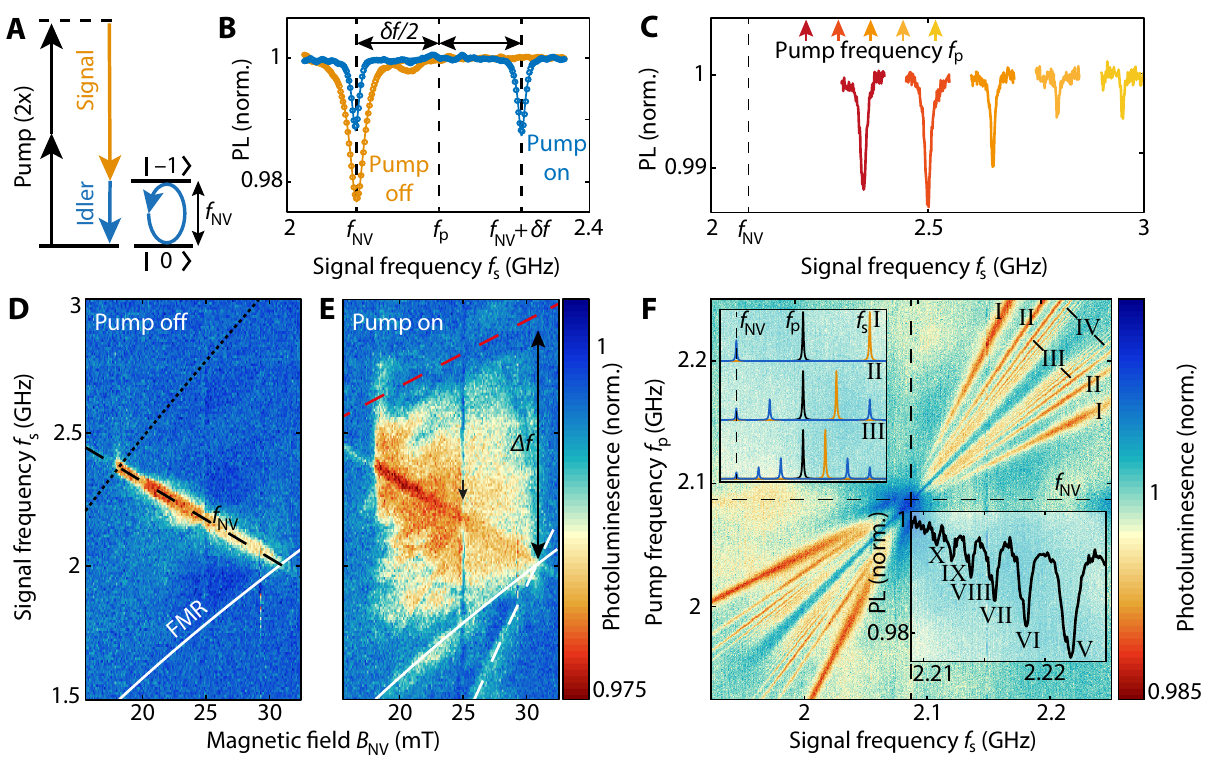}
	\caption{\textbf{Microwave detection via four-spin-wave mixing and frequency combs.} (\textbf{A}) Energy diagram of four-spin-wave mixing. The signal at frequency $f_\text{s}$ stimulates the conversion of the pump to the idler at $f_\text{NV}$. (\textbf{B}) Normalized NV photoluminescence (PL) versus $f_\text{s}$ . Without pump (orange data), an ESR dip is only observed when $f_\text{s}=f_\text{NV}$. With pump at $f_\text{p}=f_\text{NV}+\delta f/2$ (blue data), a signal at $f_\text{s}=f_\text{NV}+\delta f$ becomes detectable. (\textbf{C}) Tuning the pump (colored arrows) shifts the detectable signal frequency, observed through the shifting ESR dips (matching colors). (\textbf{D}) Normalized NV PL vs $f_\text{s}$ and magnetic field in the absence of a pump. Only signals at $f_\text{NV}$ (dashed black line) can be detected. Dotted black line: Frequency above which three-magnon scattering limits the spin-wave amplitude~\cite{Bertelli2021}. White line: Ferromagnetic resonance (FMR) frequency. (\textbf{E}) Applying a pump between $f_\text{s}$ and $f_\text{NV}$ opens a detection window from the FMR up to the second node (dashed red line) in the Fourier spectrum of the stripline field (Fig. S1). (Dashed) white line: Signal (Pump) drives FMR. Black arrow: Line of reduced contrast caused by scattering into the first perpendicular standing spin-wave mode~\cite{Bertelli2020}. (\textbf{F}) Spin-wave comb observed in the PL versus $f_\text{s}$ and $f_\text{p}$. Data is normalized (Fig. S2). Upper inset: Spectrum (sketch) illustrating the detection of idlers I-III (black: pump, orange: signal, blue: idlers). Lower inset: Linecut at $f_\text{p}=2.2\text{ GHz}$, showing idlers up to the tenth order.                 
	}
	\label{fig2}
\end{figure}
We characterize the bandwidth of the four-wave-mixing detection scheme by measuring the NV photoluminescence contrast as a function of the microwave signal frequency and magnetic bias field. As in Fig. 2B, when the pump field is switched off, we only detect signals resonant with $f_\text{NV}$  (Fig. 2D). In contrast, when the pump is switched on, a broad band of frequencies becomes detectable (Fig. 2E). The bandwidth $\Delta f$ of $\sim$ 1 GHz is limited from below by the ferromagnetic resonance (FMR), the spatially homogenous spin-wave mode below which spin waves cannot be excited in our measurement geometry, and from above by the limited efficiency of our 5-micron-wide stripline to excite high-momentum spin waves. As such, the bandwidth can be extended by using narrower striplines or magnetic coplanar waveguides~\cite{Che2020}. \\
At 14 dBm signal and pump power, consecutive mixing processes generate higher-order idler modes at discrete and equally spaced frequencies (Fig. 2F). Motivated by the success of their optical counterparts in high-precision spectrometry~\cite{Picque2019}, such “spin-wave frequency combs” are of great interest because of potential applications in microwave metrology~\cite{Wang2021,Koerner2022,Hula2021}. We use the spin-wave comb to realize sensitivity to multiple microwave frequencies by detecting the n-th order idler frequency, 
\be
f_\text{i}^{n}=(n+1)f_\text{p}-nf_\text{s}
\ee
when it is resonant with the ESR frequency (Fig. 2F, upper inset). An increasing number of idler modes appears with increasing drive power (Fig. S3), such that at large powers we resolve up to the $n=10\text{th}$ idler order (Fig. 2F, bottom inset). The shift of the idler frequency is amplified by the integer $n$ over the shift of the signal frequency (Eq. 1), leading to a $1/n$ decrease in the linewidth of the NV ESR response~\cite{Koerner2022} (Fig. 2F) and a correspondingly enhanced ability to resolve closely spaced signal frequencies. \\
In addition to enabling off-resonant quantum sensing, the idlers also provide a resource for off-resonant control of spin- or other quantum systems. The resolving of the NV’s 3-MHz hyperfine splitting in the idler-driven ESR spectrum (Fig. 3A) evidences the high coherence of the microwave field emitted by the idler spin wave, implying that the linewidth is determined by the drive rather than the spin-wave damping~\cite{Koerner2022}. This allows driving coherent NV spin rotations (Rabi oscillations) by pulsing the pump with varying duration $\tau$ (Fig. 3B). \\
Remarkably, these Rabi oscillations respond to externally applied microwaves that are detuned by hundreds of MHz from the ESR frequency (Fig. 3C). Such magnon-mediated, off-resonant Rabi control is a new instrument in the toolbox of spin-manipulation techniques, providing universal off-resonant quantum control with potential applications in quantum information processing. The idler-driven Rabi frequency exceeds the signal-induced AC Stark shift~\cite{Wei1997} by about an order of magnitude for the same off-resonant signal power (Fig. S4). The decrease of the Rabi frequency with increasing detuning $\delta f$ (Fig. 3C) is attributed to a reduced spin-wave excitation efficiency at higher frequency and a reduced frequency-conversion efficiency due to an increased momentum mismatch between signal and pump spin waves~\cite{Schultheiss2012}. \\
\begin{figure}[h!]\centering
	\includegraphics[scale=1.1]{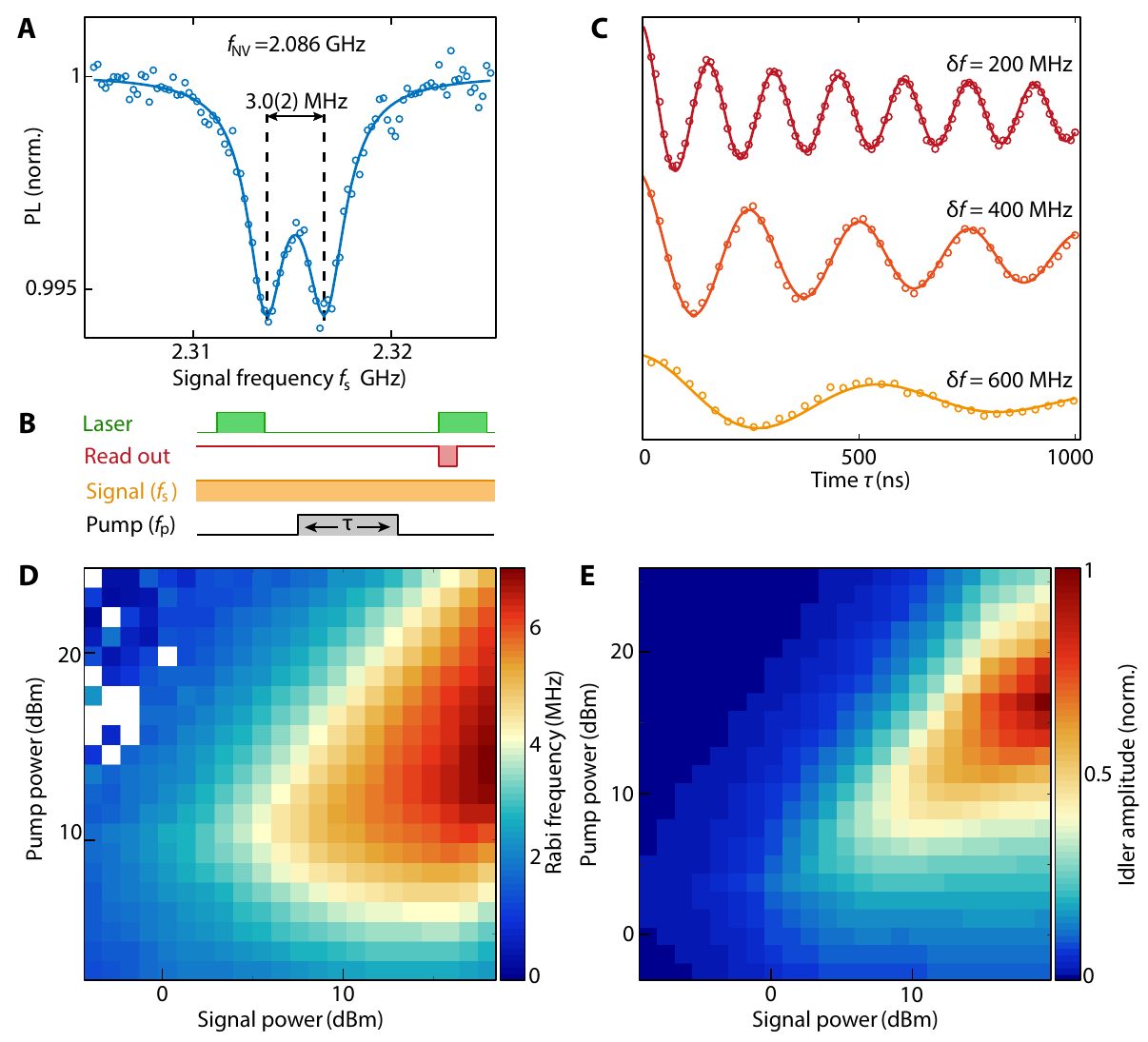}
	\caption{\textbf{Off-resonant quantum control of NV spins via frequency conversion based on four-spin-wave mixing. }(\textbf{A}) Idler-driven NV ESR spectrum with the ESR frequency at $f_\text{NV}=2.086\text{ GHz}$ and the pump at $f_\text{p}=2.2\text{ GHz}$. The narrow linewidth of the idler allows resolving the 3-MHz hyperfine splitting associated with the $^{15}$N nucleus. (\textbf{B}) Pulse sequence for driving coherent NV spin rotations (Rabi oscillations) with an off-resonant signal field. The pulsed pump and continuous-wave signal generate a pulsed idler at $f_\text{NV}$ that drives Rabi oscillations. (\textbf{C}) Optically-detected Rabi oscillations driven by the first-order idler mode for different detuning $\delta f=f_\text{s}-f_\text{NV}$. (\textbf{D}) Frequency of the idler-driven Rabi oscillations versus power of the pump ($f_\text{p}=2.2\text{ GHz}$) and signal ($f_\text{s}=2.314\text{ GHz}$). The Rabi frequency initially increases with both signal and pump power, but then decreases because of spin-wave instabilities. This non-monotonic behavior is reproduced by numerical calculations of the normalized idler amplitude in (\textbf{E}) (details in the Supplementary Information).                 
	}
	\label{fig3}
\end{figure}
Since the Rabi frequency depends linearly on the idler amplitude~\cite{Bertelli2020}, it provides insight into the magnetization dynamics in the film. As expected, the idler amplitude initially grows with increasing signal and pump power~\cite{Marsh2012,Hula2021}, but then reaches a maximum and starts to decrease (Fig. 3D). We attribute the decrease to Suhl instabilities of the second type~\cite{Suhl1957}: Both signal and pump modes decay into a pair of high-momentum magnons beyond a certain threshold amplitude, which drains energy from the idler mode. This interpretation is supported by a model of the four-wave interactions between the dominant two idler modes, the signal and pump modes, and the two pairs of high-momentum “Suhl” magnons (Figs. S5 and S6). The intermode coupling is induced by exchange and dipolar interactions, as well as crystalline anisotropy, and follows from the leading-order terms in the Holstein-Primakoff expansion~\cite{Krivosik2010}. Based on the interacting eight-mode Hamiltonian we compute the steady-state dynamics of the idler mode as a function of pump and signal power (Fig. 3E), which qualitatively reproduces the observed power dependence in Fig. 3D. Nanopatterning the magnetic film into a magnonic crystal modulates the magnon density of states, which may suppress the decay into unwanted spin-wave modes~\cite{Chumak2015,Li2019a} and thereby enable higher idler-driven Rabi frequencies.\\
\begin{figure}[h!]\centering
	\includegraphics[scale=1.3]{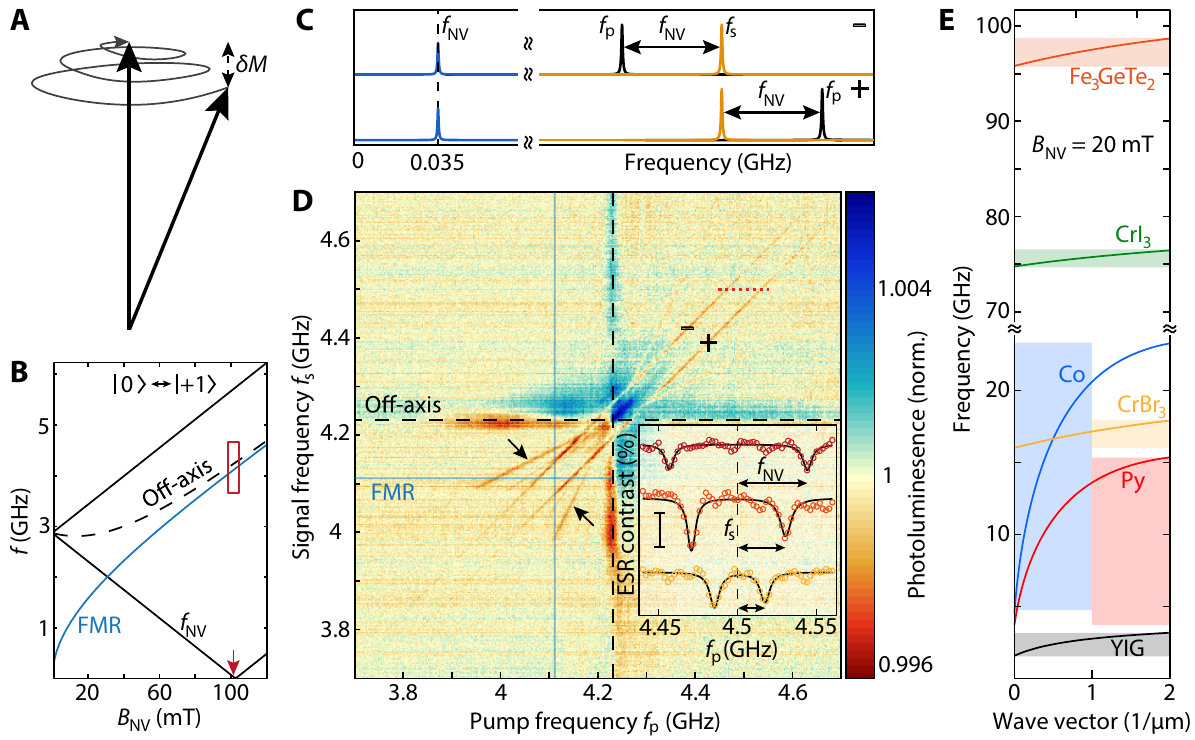}
	\caption{\textbf{Microwave detection via difference-frequency generation.} (\textbf{A}) Sketch of the spiraling precession of the magnetization $M$ when driven by microwave frequencies $f_\text{s}$ and $f_\text{p}$. The longitudinal component of the magnetization oscillates with amplitude $\delta M$ at the difference frequency ($|f_\text{s}-f_\text{p}|$), which is detected when resonant with $f_\text{NV}$. (\textbf{B}) Field dependence of the FMR (solid blue line) and the NV ESR frequencies, with the solid (dashed) black line corresponding to the on-axis (off-axis) NV families. $f_\text{NV}$ of the on-axis family enters the MHz regime near $B_\text{NV}=100\text{ mT}$ (red arrow). (\textbf{C}) Sketch of the frequency spectrum. Difference-frequency generation creates a detectable signal at $f_\text{NV}$ when the pump frequency is at $f_\text{p}=f_\text{s}\pm f_\text{NV}$. (\textbf{D}) Photoluminescence versus $f_\text{s}$  and $f_\text{p}$ at $B_\text{NV}=101.3\text{ mT}$, such that $f_\text{NV}=35\text{ MHz}$ [red box in (B)]. The parallel, diagonal lines [labeled + and – as in (C)] indicate the difference-frequency detection. Also visible are idler spin waves (indicated by the arrows) generated by four-wave mixing at the off-axis ESR frequency (dashed black lines). The data is normalized, leaving artefacts at the off-axis frequency (Methods). Inset: Line traces at $f_\text{s}=4.5\text{ GHz}$ (dotted red line in main panel) for different $B_\text{NV}$, showing the shift of the frequency difference resonant with $f_\text{NV}$. Scalebar: 0.1\% ESR contrast. (\textbf{E}) Sketched spin-wave dispersion of various 100 nm-thick magnets for $B_\text{NV}=20\text{ mT}$. Matched color shading highlights the accessible detection bandwidth using a 500-nm-wide stripline.
	}
	\label{fig4}
\end{figure}
Our second detection protocol relies on difference-frequency generation, which enables down-conversion of GHz signals to MHz frequencies accessible to established quantum sensing techniques~\cite{Degen2017}. The difference frequency is generated by the longitudinal component of the magnetization under the driving of two spin waves of different frequencies~\cite{Flebus2018}. In contrast to the four-wave mixing protocol, the converted frequency does not have to lie within the spin-wave band. By tuning the ESR frequency into resonance with the difference frequency (Fig. 4B), we detect microwave signals that are detuned by several gigahertz when $f_\text{p}-f_\text{s}=\pm f_\text{NV}$ (Fig. 4C). Alternatively, AC magnetometry protocols can provide difference-frequency detection with enhanced sensitivity at arbitrary bias ~\cite{Degen2017}. We only observe ESR contrast when both $f_\text{s}$ and $f_\text{p}$ are above the FMR (Fig. 4D), confirming that the conversion is mediated by spin waves. As such, the efficiency of the conversion process can be used to characterize high-frequency magnetic band structures. Similar to Fig. 2E, the conversion is limited by the spin-wave excitation efficiency, which explains the observation of the largest ESR contrast for long-wavelength spin waves (i.e. just above the FMR). \\
We demonstrated magnon-mediated, spin-based sensing of microwave magnetic fields over a gigahertz bandwidth at a fixed magnetic bias field. The frequency of the pump determines the detection frequency, and the detection range is set by the frequencies at which spin waves can be excited efficiently. The range could be extended to the 10-100 GHz scale using materials with a larger magnetization that increases the spin-wave group velocity or crystal anisotropies that increase the spin-wave gap (Fig. 4E). The coherent nature of the frequency conversion allows combining with advanced spin-manipulation protocols such as heterodyne or dressed-state sensing~\cite{Meinel2021,Joas2017,Stark2017} to further increase the detection capabilities. Wide-field readout of NV centers in a larger sensing volume would enhance the sensitivity, which is ultimately limited by thermally generated spin-wave noise. The demonstrated hybrid diamond-magnet sensor platform enables broadband microwave characterization without requiring large magnetic bias fields and opens the way for probing high-frequency magnetic spectra of new materials, such as van-der-Waals magnets.
\section*{Methods}
\textbf{Experimental setup.} The NV photoluminescence is read out using a confocal microscope described in Ref.~\cite{Bertelli2020}. The NV-YIG chip and its fabrication were described in Ref.~\cite{Bertelli2021}. It consists of a 2x2x0.05-mm$^3$ diamond membrane with an estimated near-surface NV density of $10^3/\upmu\text{m}^2$ placed on top of a 235-nm-thick YIG film grown using liquid phase epitaxy on a 500-$\upmu$m-thick GGG substrate (Matesy GmbH). The diamond-YIG separation distance is $\sim$ 2 $\upmu$m, limited by small particles (such as dust) between the diamond and the YIG surfaces. The signal and pump microwaves are generated by two Rohde $\&$ Schwarz microwave sources (SGS100A), combined by a Mini-Circuits power combiner (ZFRSC-123-S+, total loss: $\sim$ -10 dB) and amplified by an AR amplifier (30S1G6, amplification: $\sim$ 44 dB). All measurements were performed at room temperature. \\

\textbf{NV microwave magnetometry.} The four NV-center families are sensitive to microwave magnetic fields at their electron spin resonance (ESR) frequencies, which are determined by the magnetic bias field $\textbf{B}_\text{NV}$ via the NV spin Hamiltonian $H=DS_z^2+\gamma\textbf{B}_\text{NV}\cdot\textbf{S}$, with $D=2.87\text{ GHz}$ the zero-field splitting, $\gamma=28\text{ GHz/T}$ the electron gyromagnetic ratio and $S_{i\in\{x,y,z\}}$ the $i$th spin-1 Pauli matrix. In this work, we align the field along one of the NV orientations, such that this “on-axis” family has $\ket{0}\leftrightarrow\ket{\pm1}$ ESR frequencies given by $D\pm\gamma B_\text{NV}$  (with $B_\text{NV}=|\textbf{B}_\text{NV}|$). For the other three “off-axis” families, the bias field is equally misaligned by $\sim71^\circ$ due to crystal symmetry, leading to the ESR frequency plotted in Fig. 4B (labeled “Off-axis”). The photoluminescence dips were recorded using continuous-wave microwaves and non-resonant optical excitation at 515 nm. For the Rabi oscillations we first initialize the NV spin in the $\ket{0}$-state via a $\sim$ 1-$\upmu$s green laser pulse, then we drive the spin using an idler pulse and finally we read out the NV photons in the first 300-400 ns of a second laser pulse. \\

\textbf{Data processing.} The data presented in Figs. 2F and 4D is normalized by the median of each row and column (Fig. S2).
\printbibliography
\section*{Acknowledgments}
We thank M. N. Ali for reviewing the manuscript. This work was supported by the Dutch Research Council (NWO) through the Frontiers of Nanoscience (NanoFront) program, the NWO Projectruimte grant 680.91.115, the Kavli Institute of Nanoscience Delft, and the Japan Society for the Promotion of Science (JSPS) by Kakenhi Grant $\#$ 19H00645.\\
\textbf{Author contributions.} J.J.C. and T.v.d.S. conceived the experiment. I.B. built the setup and fabricated the sample. R.W.M., J.J.C., I.B. and A.T. performed the measurements and analyzed the data. M.E., Y.M.B. and G.E.W.B. developed the theoretical model for the idler amplitude, B.G.S. fabricated the diamond membrane, J.J.C., T.v.d.S., I.B. and A.T. wrote the manuscript with help from all coauthors. \\
\textbf{Competing interests.} The authors declare that they have no competing interests. \\
\textbf{Data availability.} The numerical data plotted in the figures in this work are available at Zenodo with identifier \url{https://doi.org/10.5281/zenodo.6543615}. The codes used for the numerical calculation of the idler amplitude are available upon request. 
\end{refsection}
\newpage
\beginsupplement 
\setstretch{1.25}
\begin{refsection} 
\begin{center}
{\LARGE Supplementary Information}\break\break\break
\textbf{{\Large\textbf{Broadband microwave detection using electron spins in a hybrid diamond-magnet sensor chip }}}\break\break\break
{\normalsize Joris J. Carmiggelt}$^{1}$, {\normalsize Iacopo Bertelli}$^{1}$, {\normalsize Roland W. Mulder}$^{1}$, {\normalsize Annick Teepe}$^{1}$, {\normalsize Mehrdad Elyasi}$^{2}$, \break  {\normalsize Brecht G. Simon}$^{1}$, {\normalsize Gerrit E. W. Bauer}$^{1,2,3}$, {\normalsize Yaroslav M. Blanter}$^{1}$, {\normalsize Toeno van der Sar}$^{1,*}$\break
\end{center}
\begin{flushleft}
\footnotesize{
\hspace*{14pt}$^1$Department of Quantum Nanoscience, Kavli Institute of Nanoscience, Delft University of Technology; \\ \raggedright \parindent=15pt Lorentzweg 1, 2628 CJ Delft, The Netherlands. \\
$^2$Advanced institute for Materials Research (WPI-AIMR), Tohoku University; Sendai 980-8577, Japan.\\
$^3$Kavli Institute for Theoretical Sciences, University of Chinese Academy of Sciences; Beijing 100190, China. \\
}
\end{flushleft}
\begin{center}
\footnotesize{
$^*$Corresponding author. Email: \href{mailto:T.vanderSar@tudelft.nl}{T.vanderSar@tudelft.nl}
}
\end{center}

\phantomsection
\addcontentsline{toc}{section}{Supplementary Information}
\newpage
\subsection{Derivation of the spin-wave dispersion for bias fields along the NV axis}\label{Sup:Disp}
Here we derive the spin-wave dispersion for a magnetic film in the $xy$-plane with perpendicular magnetic anisotropy (PMA) and a magnetic bias field $\textbf{B}_B$ in an arbitrary direction. The dispersion is given by the poles of the transverse magnetic susceptibility~\cite{Bertelli2021, Rustagi2020} that relates the transverse magnetization to a drive field $\textbf{B}_\text{AC}$. We derive the magnetic susceptibility from the Landau-Lifshitz-Gilbert (LLG) equation that describes the dynamics of the unit magnetization vector $\textbf{m}$
\be
\dot{\bfm} = -\gamma \bfm\times \bfB-\alpha_\text{G} \dot{\bfm}\times \bfm,
\label{LLG}
\ee	
where $\alpha_\text{G}$ is the Gilbert damping and the “overdot” denotes the time derivative. We solve this equation in the ($x'$,$y'$,$z'$) magnet frame that is tilted with respect to the ($x$,$y$,$z$) lab frame by an angle $\theta_0$, such that the equilibrium magnetization points in the $\uz'$ direction and the $\uy^{(')}$ axes overlap. $\textbf{B}=\bfB_\text{eff}+ \bfB_\text{AC}$, with $\bfB_\text{eff}$ the effective magnetic field as derivative of the magnetic free energy density $F$
\be 
B_{\text{eff},\alpha'}= -\frac{1}{M_\text s}\frac{\partial F}{\partial m_{\alpha'}},
\label{Beff}
\ee 
where $M_\text s$ is the saturation magnetization and $\alpha'\in\{x',y',z'\}$ indicates the vector components in the magnet frame. The free energy density includes the Zeeman energy, the demagnetizing field $\textbf{B}_\text{d}$, the PMA energy $F_\text{A}$, and the exchange interaction
\be
F = -M_\text s\mathbf{m}\cdot(\textbf{B}_B+\frac{\mathbf{B}_\text{d}}{2})+F_\text{A}+ \frac{D}{2}
\sum_{\alpha,\beta=x,y,z}\left(\frac{\partial m_{\alpha'}}{\partial \beta}\right)^2.
\label{FF}
\ee
In the magnet frame
\be
F_\text{A}=\frac{K}{2}m_z^2=\frac{K}{2}(\sin\theta_0m_{x'}+\cos\theta_0m_{z'})^2,
\ee
such that the $x'$ and $z'$ components of the anisotropy effective field are
\be
\begin{split}
&B_{\text{A,}x'}=-\frac{1}{M_\text s}\frac{\partial F}{\partial m_{x'}}=-\frac{K}{M_\text s}(\sin^2\theta_0m_{x'}+\cos\theta_0\sin\theta_0m_{z'}), \\
&B_{\text{A,}z'}=-\frac{1}{M_\text s}\frac{\partial F}{\partial m_{z'}}=-\frac{K}{M_\text s}(\cos\theta_0\sin\theta_0m_{x'}+\cos^2\theta_0m_{z'}).
\end{split}
\ee
The contributions of the Zeeman-, demagnetizing- and exchange energy to $\mathbf{B}_\text{eff}$ have been derived in Refs.~\cite{Bertelli2021, Rustagi2020}. \\
In linear response with $m_{z'}\approx1$,  the LLG equation describes the transverse magnetization dynamics. In the frequency domain it reads
\be
\begin{split}
&-i\omega m_{x'}=-\gamma(m_{y'}B_{z'}-B_{y'})+i\alpha_\text{G}\omega m_{y'},\\
&-i\omega m_{y'}=\gamma(m_{x'}B_{z'}-B_{x'})-i\alpha_\text{G}\omega m_{x'},
\end{split}
\ee
where $\omega$ is the angular frequency. Substituting the components of the effective magnetic field and rewriting the equations in matrix form, 
\be
\begin{pmatrix}
\omega_2-i\alpha_\text{G}\omega & -\omega_1+i\omega \\
-\omega_1-i\omega& \omega_3-i\alpha_\text{G}\omega
\end{pmatrix}
\begin{pmatrix}
m_{x'} \\
m_{y'}
\end{pmatrix}
=\gamma
\begin{pmatrix}
B_{\text{AC},x'} \\
B_{\text{AC},y'}
\end{pmatrix},
\label{needinv}
\ee
where
\be
\begin{split}
&\omega_0=-(\omega_M-\omega_K)\cos^2\theta_0+\omega_B\cos(\theta_B-\theta_0)+\omega_Dk^2, \\
&\omega_1=\omega_Mf\sin\phi\cos\phi\cos\theta_0,\\
&\omega_2=\omega_0+\omega_Mf(\cos^2\phi\cos^2\theta_0-\sin^2\theta_0)+(\omega_M-\omega_K)\sin^2\theta_0,\\
&\omega_3=\omega_0+\omega_Mf\sin^2\phi,
\end{split}
\ee
and $\omega_B=\gamma B_B$, $\omega_M=\gamma\mu_0M_\text{s}$, $\omega_D=\frac{\gamma D}{M_\text{s}}$ and $\omega_K=\frac{\gamma K}{M_\text{s}}$. $\mu_0$ is the vacuum permeability, $k=|\textbf{k}|$ is the modulus of the wavevector along an angle $\phi$ with respect to the in-plane projection of the magnetization, $\theta_B$ is the angle of the magnetic bias field with respect to the plane normal ($z$ axis), and $f=1-\frac{1-e^{-kt}}{kt}$ depends on the film thickness $t$. By inverting the matrix in Eq. (S7), we obtain the transverse magnetic susceptibility, which is singular when 
\be
(\omega_2-i\alpha_\text{G}\omega)(\omega_3-i\alpha_\text{G}\omega)-\omega_1^2-\omega^2=0.
\ee
Assuming $\alpha_\text{G}\ll1$, the real part of the solutions of this quadratic equation gives the spin-wave dispersion as a function of $k$
\be
\omega^2=\omega_2\omega_3-\omega_1^2.
\ee
The theoretical lines in Figs. 2 and 4 in the main text are based on Eq. (S10). We assume that the field is applied parallel to the NV axis, such that $\theta_B=54.7^\circ$, with in-plane projection along the stripline. We consider only spin waves with $\phi=\pi/2$, since these are most efficiently excited by our 150-micron-long stripline. $\theta_0$ minimizes the free energy density and is found by numerically solving $\frac{\partial F}{\partial\theta_0}=0$. The ferromagnetic resonance (FMR) frequency corresponds to $k=0$. Table S1 states the values of the saturation magnetization, exchange and uniaxial anisotropy constants for different magnetic materials used for calculating the spin-wave dispersions in Fig. 4E of the main text.

\subsection{Dependence of the detection bandwidth on the microwave drive field}\label{Sup:Drive}
For efficient frequency conversion, the microwaves should excite propagating spin waves with a significant amplitude. The spin-wave excitation efficiency depends on the microwave power and the spatial mode overlap between the drive field and the spin waves~\cite{Bertelli2020}. In our experiment, a 5-micron-wide stripline creates an inhomogeneous microwave drive field with a sinc-like amplitude in $k$-space (Fig. S1A). The efficiency drops with decreasing wavelength with nodes at $\lambda=w/n$, where $w$ is the stripline width and $n$ is an integer.\\
To characterize the dependence on the drive field, we measure the bandwidth induced by four-wave mixing as a function of the pump power (Fig. S1B). As expected, the bandwidth increases with microwave power. The photoluminescence contrast is suppressed at spin-wave frequencies that correspond to the nodes of the drive field in Fig. S1A (colored dashed lines). The frequencies of these modes agree with the spin-wave dispersion derived in the previous section [Eq. (S10)]. The spin-wave excitation antenna is therefore an important design parameter for hybrid diamond-magnet microwave sensors. 

\subsection{Comparison between the idler-driven Rabi frequency and dynamical Stark shift}\label{Sup:Stark}
A strong microwave field detuned by $\delta f$ from the NV ESR frequency ($f_\text{NV}$), causes the latter to shift, an effect known as the AC (or dynamical) Stark shift~\cite{Wei1997}. The Stark shift increases with drive power and is inversely proportional to $\delta f$, which allows detecting the presence of an off-resonant microwave signal. We show here that the idler-driven Rabi frequency resulting from four-spin-wave mixing is about an order magnitude larger than the Stark shift at the same off-resonant drive power.\\ 
We measure the Stark shift via pump-probe microwave spectroscopy. The high-power pump is detuned from $f_\text{NV}$ by 10-1000 MHz, while a low-power probe measures the ESR frequency. We determine the Stark shift for every detuning by measuring the ESR frequency with and without pump (blue data in Fig. S4A). \\
Next, we measure Rabi oscillations using the four-spin-wave mixing technique. We extract the Rabi frequency for signal spin waves detuned from 10 to 710 MHz (red data in Fig. S4A). We attribute the small oscillations in the Rabi frequency and Stark shift to frequency-dependent (cable) resonances in the microwave transmission of the stripline. Fig. S4B shows that the Rabi frequencies are larger than the Stark shift by about an order of magnitude over the measurement range. 

\subsection{Eight-modes model}\label{Sup:Model}
Here we describe the details of the model for the spin-wave dynamics under a two-tone drive used to calculate the idler amplitude as a function of pump and signal power, as plotted in Fig. 3E in the main text. \\
Fig. S5 shows the spin-wave dispersion of a YIG film for $\theta_\bfk=0$ (blue line) and $\theta_\bfk=\pi/2$ (black line), where $\theta_\bfk$ is the angle between the in-plane spin-wave wavevector $\bfk$ and the static magnetization for the parameters in Table S1. Since the out-of-plane component of the applied bias field $\bfB_\text{NV}$ is small compared to the demagnetizing field of $\sim$ 178 mT in YIG, we assume that the static magnetization lies in-plane along $\uz$ ($\ux$ is the out-of-plane axis), parallel to $B_\uz$, the in-plane component of $\bfB_\text{NV}$. The long stripline along $\uz$ excites signal and pump spin waves with $\theta_\bfk=\pi/2$. Conservation of momentum dictates that the two created idler spin waves also lie on the $\theta_\bfk=\pi/2$ branch with wavevectors $\bfk_\text i=2\bfk_\text p-\bfk_\text s$ and $\bfk_{\text{i}'}=2\bfk_\text s-\bfk_\text p$ (Fig. S5). \\
When the pump mode is strongly driven beyond a certain threshold, the four magnon scattering term in the spin-wave Hamiltonian  $c_{\bfk_{\text{p}}}^{\dag}c_{\bfk_{\text{p}}}^{\dag}c_{\bfk^{\prime}}c_{\bfk^{\prime\prime}}$ leads to a Suhl instability. Here $c_{\bfk}^{(\dag)}$ is the annihilation (creation) operator for a magnon with wavevector $\bfk$, which is normalized by $\sqrt{S}$, where $S=Vs_n/V_n$ is the total number of spins, $V$ is the volume, $s_n$ is the number of spins per unit cell, and $V_n$ is the unit cell volume. A specific pair of magnons wins the “instability competition”, $\bfk'=\bfk_{\text p, 1}$ and $\bfk''=\bfk_{\text p, 2}=2\bfk_{\text p}-\bfk_{\text p, 1}$, which we call the “efficient Suhl pair” of the pump mode. The efficient Suhl pair for the signal $\bfk_{\text s, 1}$ and $\bfk_{\text s, 2}$ should also be considered when its mode amplitude is sufficiently large. We disregard cascades that lead to the weak higher-order idlers in Fig. 2F of the main text, as well as the Suhl pairs of the idlers that are safely below their instability threshold at the presently applied powers. A minimal model should therefore include the eight modes indicated in Fig. S5. \\
The efficient pump and signal pairs can be identified from the threshold amplitude of the pump (signal) mode $x=|\alpha_{\text p, \text s}|^2$ above which the Suhl instability leads to $\{\bfk',\bfk''\}$ pairs, which solve~\cite{Elyasi2022}
\be
({{\mathcal{D}}^{\text{Suhl}}_{\text{p}(\text{s});\bfk';\bfk''}}^2-{{\mathcal{D}}^{\text{CK}}_{\text{p}(\text{s});\bfk'}}^{2})x^{2}-2{\mathrm{\Delta}}{\mathcal{D}}^{\text{CK}}_{\text{p}(\text{s});\bfk'}x-{\xi}^{2}-{\mathrm{\Delta}}^{2}=0,
\ee
where  $\Delta=\omega_{\text{p}(\text{s})}-(\omega_{\bfk'}+\omega_{\bfk''})/2$, with $\omega$ an angular frequency, and $\xi$ is a dissipation rate chosen here to be 10 MHz for all modes.  ${{\mathcal{D}}^{\text{Suhl}}_{\text{p}(\text{s});\bfk';\bfk''}}$ (${{\mathcal{D}}^{\text{CK}}_{\text{p}(\text{s});\bfk'}}$) is the matrix element for the scattering process  $c_{\bfk_{\text{p}(\text{s})}}^{\dag}c_{\bfk_{\text{p}(\text{s})}}^{\dag}c_{\bfk^{\prime}}c_{\bfk^{\prime\prime}}$ ($c_{\bfk_{\text{p}(\text{s})}}^{\dag}c_{\bfk_{\text{p}(\text{s})}}c_{\bfk^{\prime}}^{\dag}c_{\bfk^{\prime}}$) in the Hamiltonian~\cite{Krivosik2010}. First, we numerically calculate the threshold amplitude $|\alpha_{\text p, \text s}|^2$ as a function of $|\bfk'|$ and $\theta_{\bfk'}$ as in Fig. S6A. We identify the minimum threshold amplitude in the ($|\bfk'|$,$\theta_{\bfk'}$) plane of Fig. S6A as a function of modulus $|\bfk'|$ in Fig. S6B. The corresponding $\theta_{\bfk'}$ and spin-wave pair frequencies are shown in Figs. S6B and S6C, respectively. The spin-wave pair with the lowest threshold amplitude –  the effective pump (signal) Suhl pair – turns out to be at angles $\theta_{\bfk'}$ far from $\pi/2$ (as indicated by the vertical dashed lines in Fig. S6B-C). \\
Our model Hamiltonian reads
\be
\begin{split}
&H=\sum_{X}\omega_{X}c^{\dag}_{X}c_{X}+[\mathcal{D}^{\text{Suhl}}_{\text{p};\text{p},1;\text{p},2}c_{\bfk_{\text{p}}}^{\dag}c_{\bfk_{\text{p}}}^{\dag}c_{\bfk_{\text{p},1}}c_{\bfk_{\text{p},2}}+\\
&\mathcal{D}^{\text{Suhl}}_{\text{p};\text{s};\text{i}}c_{\bfk_{\text{p}}}^{\dag}c_{\bfk_{\text{p}}}^{\dag}c_{\bfk_{\text{s}}}c_{\bfk_{\text{i}}}+\mathcal{D}^{\text{Suhl}}_{\text{s};\text{s},1;\text{s},2}c_{\bfk_{\text{s}}}^{\dag}c_{\bfk_{\text{s}}}^{\dag}c_{\bfk_{\text{s},1}}c_{\bfk_{\text{s},2}}+\\
&\mathcal{D}^{\text{Suhl}}_{\text{s};\text{p};\text{i}^{\prime}}c_{\bfk_{\text{s}}}^{\dag}c_{\bfk_{\text{s}}}^{\dag}c_{\bfk_{\text{p}}}c_{\bfk_{\text{i}^{\prime}}}+\text{H.c.}]+\\
&\sum_{X}[\mathcal{D}^{\text{SK}}_{X}c_{X}^{\dag}c_{X}c_{X}^{\dag}c_{X}+\sum_{Y}\mathcal{D}^{\text{CK}}_{X;Y}c_{X}^{\dag}c_{X}c_{Y}^{\dag}c_{Y}]+\\
&E'_{\text{p}}(e^{-i\omega_{\text{p}}t}c_{\text{p}}^{\dag}+\text{H.c.})+E'_{\text{s}}(e^{-i\omega_{\text{s}}t}c_{\text{s}}^{\dag}+\text{H.c.}).
\end{split}
\ee
Here $X,Y\in\{\text p;\text s;\text i;\text i';\text p, 1(2);\text s, 1(2)\}$, and “H.c.” denotes the Hermitian conjugate. $E'_\text s$ and $E'_\text p$ are the drive amplitudes of the signal and pump modes, respectively, which are related to the excitation power of the microstrip by (cf. Fig. S1A), 
\be
P_{\text{p}(\text{s})}={E_{\text{p}(\text{s})}^2Z(\omega_{\text{p}(\text{s})})}\left[\frac{\mu_{0}\gamma\left(e^{-\vert \bfk_{\text{p}(\text{s})}\vert d}-1\right)\sin\left(\frac{\vert\bfk_{\text{p}(\text{s})}\vert W}{2}\right)}{Wd\vert \bfk_{p(s)}\vert^2}\right]^{-2}.
\ee
Here, $\mu_0$ is the vacuum permeability, $d$ is the thickness of the stripline and $W$ is its width, $Z(\omega)$ is the impedance at $\omega$, and we assumed $\theta_{\bfk_{\text p(\text s)}}=\pi/2$. We adopt $Z(\omega_{\text p(\text s)})=50$ $\Omega$, $d=200$ nm, $W=5$ $\upmu$m, $E_{\text p(\text s)}=E'_{\text p(\text s)}\sqrt{L}$, where $L\sim W$ is the length of the excited part of the sample, $\omega_\text{p}/2\pi=2.2\text{ GHz}$, $\omega_\text{s}/2\pi=2.32\text{ GHz}$ and $B_\uz=23\text{ mT}$, corresponding to $B_\text{NV}\sim 28\text{ mT}$. From the four-magnon-scattering parameters  $\mathcal{D}^{\text{Suhl}}_{\text{p};\text{p},1;\text{p},2}/\mathcal{D}^{\text{Suhl}}_{\text{p};\text{s};\text{i}}\sim\mathcal{D}^{\text{Suhl}}_{\text{s};\text{s},1;\text{s},2}/\mathcal{D}^{\text{Suhl}}_{\text{s};\text{p};\text{i}^{\prime}}\sim 10$, and $\mathcal{D}^{\text{Suhl}}_{\text{s};\text{p};\text{i}^{\prime}}\sim \mathcal{D}^{\text{Suhl}}_{\text{p};\text{s};\text{i}}=-7.2\text{ GHz}$, and $\mathcal{D}^{\text{SK}}_{\text{p}(\text{s},\text{i},\text{i}^{\prime})}\sim\mathcal{D}^{\text{CK}}_{\text{p}(\text{s},\text{i},\text{i}^{\prime}),\text{p}(\text{s},\text{i},\text{i}^{\prime})}\sim\mathcal{D}^{\text{Suhl}}_{\text{p};\text{s};\text{i}}$, and $\mathcal{D}^{\text{SK}}_{\text{s},1(\text{s},2;\text{p},1;\text{p},2)}\sim\mathcal{D}^{\text{CK}}_{\text{p}(\text{s}),\text{s},1(\text{s},2;\text{p},1;\text{p},2)}\sim\mathcal{D}^{\text{Suhl}}_{\text{p};\text{p},1;\text{p},2}$ we calculate the mean field amplitude of the idler mode $\langle c_{\bfk_{\text{i}}}^{\dag}c_{\bfk_{\text{i}}}\rangle=\vert\alpha_{\text{i}}\vert^2$ as a function of  $P_\text s$  and $P_\text p$. In Fig. 3E of the main text we plot $\vert\alpha_{\text{i}}\vert$, since it is linearly proportional to the idler-driven Rabi frequency of the NV center~\cite{Bertelli2020}.\\
We find an idler amplitude (and thus a Rabi frequency) that initially grows as a function of pump and signal power. However, above the Suhl instability thresholds, the amplitude of the idler mode decreases due to the newly opened dissipation channels, as observed in the experiments in Fig. 3D of the main text. Since $\langle c_{X}^{\dag}c_{X}\rangle\propto E^2_{\text p(\text s)}/\xi^2$, $\xi$ can be scaled by $q$ to achieve the same phase diagram for $P_{\text p (\text s)}$ shifted by $20\cdot\log{(q)}$ dBm. Our current assumption of $\xi=10$ MHz for $\omega_X/2\pi\sim2\text{ GHz}$ corresponds to a Gilbert damping of $\alpha_\text G=5\cdot10^{-3}$.

\subsection{Difference-frequency generation by the longitudinal component of the magnetization}\label{Sup:Diff}
In this section we demonstrate that simultaneous transverse magnetization dynamics at the signal and pump frequencies ($f_\text s$  and $f_\text p$, respectively) causes a beating in the longitudinal component at the difference frequency $|f_\text s-f_\text p\vert$. The normalized transverse magnetization $m_\text T$ of two propagating circularly-polarized spin waves is the superposition
\be
m_\text T=m_\text s e^{i(k_\text s x-\omega_\text s t)}+m_\text p e^{i(k_\text p x-\omega_\text p t)}.
\ee
$k_i=2\pi/\lambda_i$ is the wavevector of the $i$th spin wave, with $i\in\{\text s, \text p\}$, in terms of the wavelength $\lambda_i$, $\omega_i=2\pi f_i$ is the angular frequency and $m_i=M_i/M_\text s$ is the magnetization amplitude normalized by the saturation magnetization. The transverse $x$ and $y$ components are the real and imaginary parts of $m_\text T$ while the normalized longitudinal component of the magnetization reads
\be
m_\text L=\sqrt{1-|m_\text T|^2}.
\ee
When driving two spin waves at frequencies $\omega_\text s$ and $\omega_\text p$, and amplitudes $m_\text s$  and $m_\text p$, the squared modulus
\be
|m_\text T|^2=m_\text Tm_\text T^*=m_\text s^2+m_\text p^2+2m_\text sm_\text p\cos((k_\text s-k_\text p)x-(\omega_\text s-\omega_\text p)t)
\ee
depends on time. For $m_i\ll1$ the longitudinal component oscillates at the difference frequency
\be
m_\text L\propto m_\text sm_\text p \cos((k_\text s-k_\text p)x-(\omega_\text s-\omega_\text p)t),
\ee
as detected in our experiments.
\newpage
\begin{figure}[h!]\centering
	\includegraphics[width=1\textwidth]{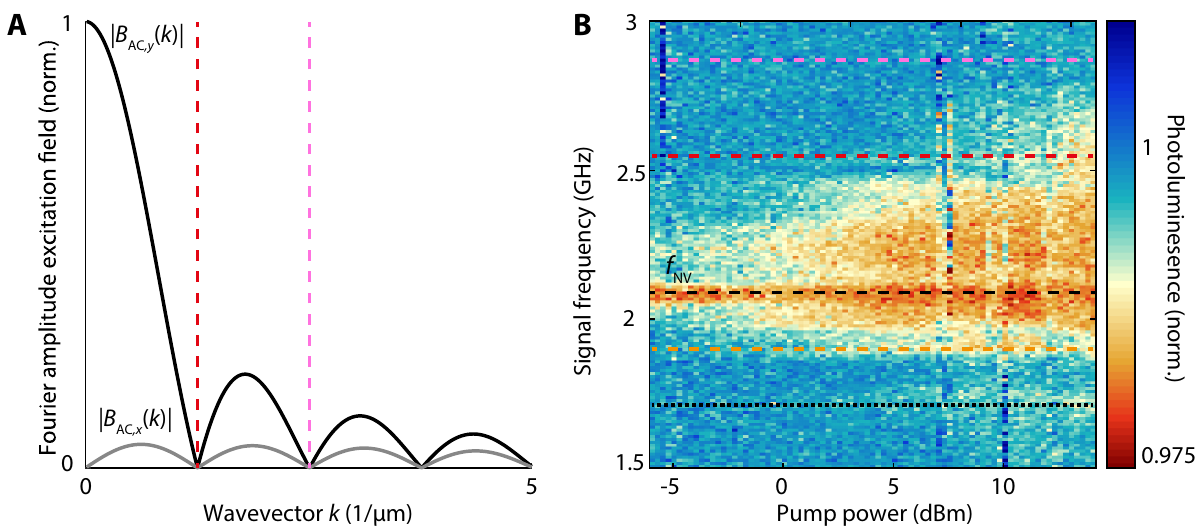}
	\caption{\textbf{The detection bandwidth is determined by the spin-wave excitation efficiency.} (\textbf{A}) Normalized Fourier amplitude of the out-of-plane ($x$, gray) and in-plane ($y$, black) components of the microwave drive field $\bfB_\text{AC}$ generated by a 5-micron-wide stripline. The colored dashed lines indicate the first two nodes in the spectrum. (\textbf{B}) Normalized NV photoluminescence induced by four-wave mixing as a function of signal frequency and pump power at a static magnetic field of $B_\text{NV}=28\text{ mT}$. The ESR frequency is at $f_\text{NV}=2.08\text{ GHz}$ (black dashed line, labeled $f_\text{NV}$) and the dashed (dotted) orange (black) lines indicate the frequencies at which the signal (pump) spin waves are driving the FMR. The red and pink horizontal dashed lines indicate the frequencies of the spin waves that nominally cannot be excited by the stripline, where colors match the nodes in (A).                
	}
	\label{figs1}
\end{figure}
\newpage 
\begin{figure}[h!]\centering
	\includegraphics[width=1\textwidth]{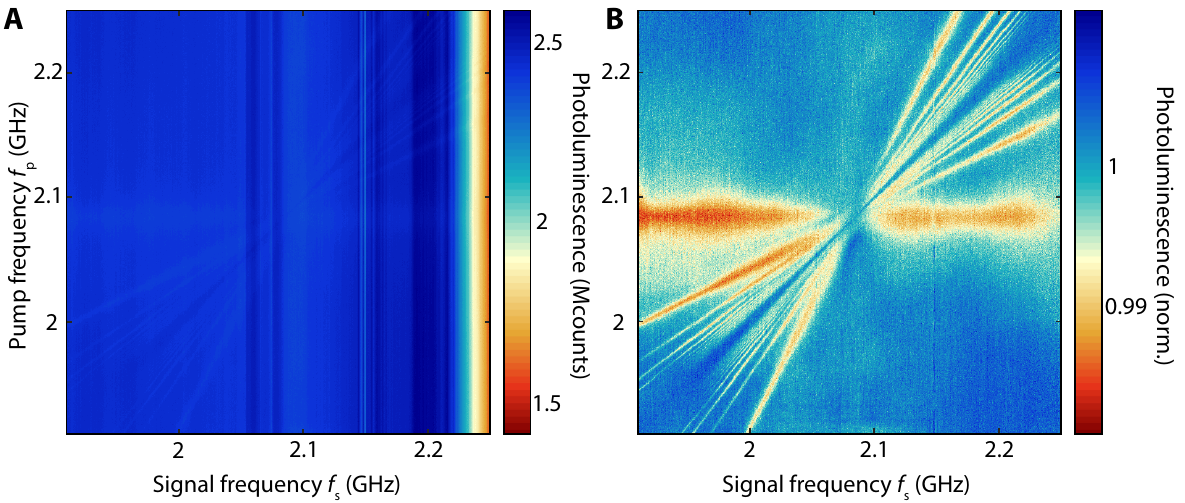}
	\caption{\textbf{Normalization procedure applied for Figs. 2F and 4D of the main text.} (\textbf{A}) Raw photoluminescence data of the measurement. We attribute the fluctuations between columns to drifts of the objective focus and laser power over the course of the measurement. (\textbf{B}) By dividing the data by the median of each column the spin-wave comb is revealed. To remove the horizontal line of photoluminescence contrast caused by resonant driving at the ESR frequency $f_\text{NV}=2.086\text{ GHz}$, we divide the data a second time by the median of each row, resulting in Fig. 2F in the main text. The same normalization procedure was applied for Fig. 4D in the main text.              
	}
	\label{figs2}
\end{figure}
\newpage 
\begin{figure}[h!]\centering
	\includegraphics[width=1\textwidth]{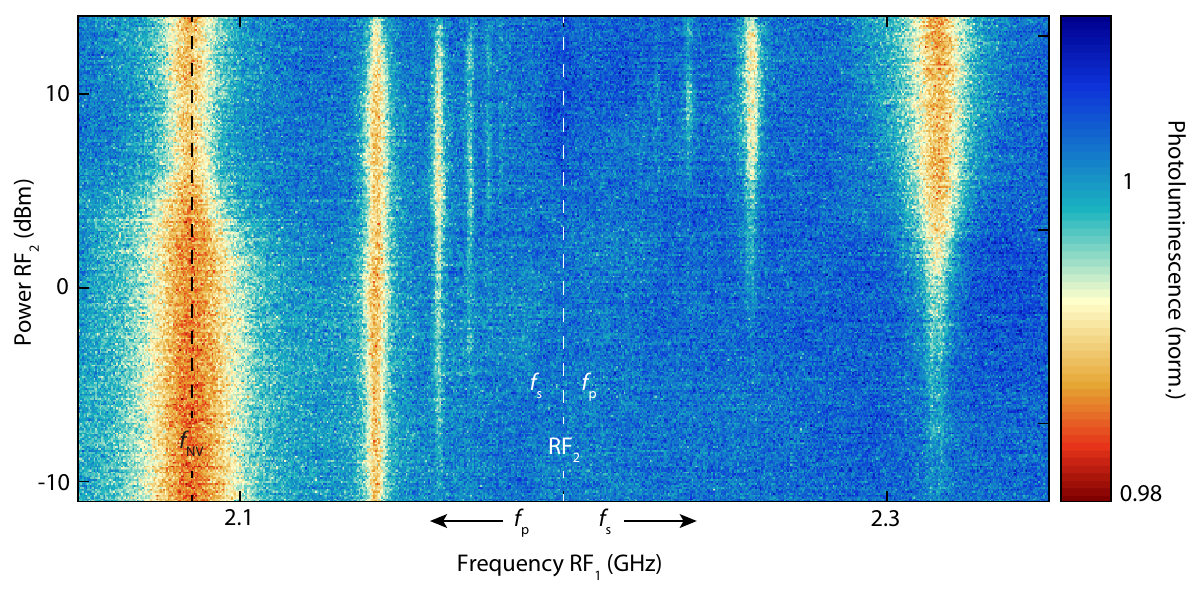}
	\caption{\textbf{Emergence of a spin-wave comb generated by two microwave drives RF$_1$ and RF$_2$.} Normalized NV photoluminescence at $B_\text{NV}=28\text{ mT}$ as a function of RF$_1$ frequency (RF$_2$ is kept at 2.2 GHz, red dashed line), and RF$_2$ power (RF$_1$ is kept at 4 dBm). An increasing number of higher-order idlers appear at increased drive power. RF$_1$ and RF$_2$ function either as pump or signal field depending on which frequency is closer to the ESR frequency $f_\text{NV}=2.086\text{ GHz}$, as is indicated by the labels $f_\text{s}$ and $f_\text{p}$ with matching colors.             
	}
	\label{figs3}
\end{figure}
\newpage 
\begin{figure}[h!]\centering
	\includegraphics[width=1\textwidth]{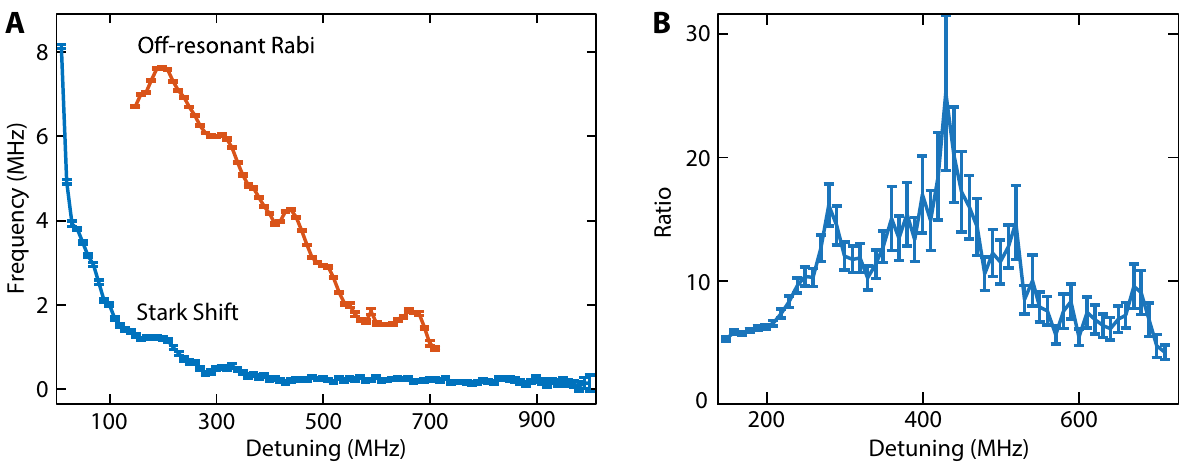}
	\caption{\textbf{Comparison of the four-spin-wave mixing and Stark shift off-resonant detection techniques.} (\textbf{A}) Blue: Measured shift in NV ESR frequency due to the AC Stark effect as a function of frequency detuning of the applied drive field. Red: Frequency of the Rabi oscillations driven by the first-order idler mode using the four-spin-wave down conversion technique as a function of drive-field detuning. (\textbf{B}) Ratio between the Rabi frequency and Stark shift as a function of detuning. The measurements were carried out at a magnetic bias field of $B_\text{NV}=28\text{ mT}$.             
	}
	\label{figs4}
\end{figure}
\newpage
\begin{figure}[h!]\centering
	\includegraphics[scale=1.5]{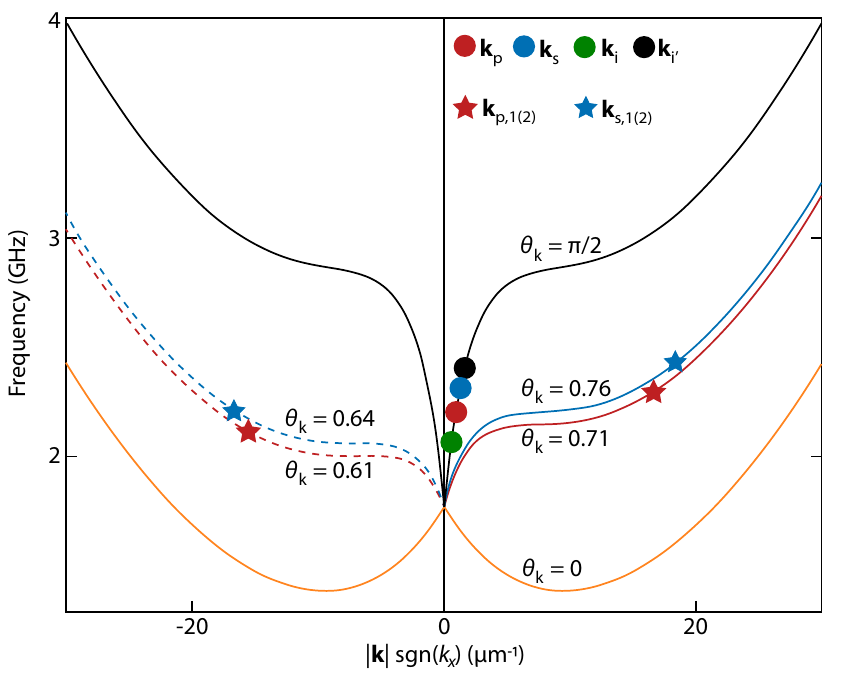}
	\caption{\textbf{Eight-modes model for the calculation of the idler amplitude.} In our model we consider the signal (blue dot), pump (red dot), idlers (green and black dots), “efficient signal Suhl instability pair” (blue stars) and “efficient pump Suhl instability pair” (red stars) spin waves. The lines are branches of the spin-wave dispersion corresponding to different angles $\theta_\bfk$ of the wavevector with respect to the static magnetization (see legend). The dispersion is symmetric upon rotations of $\theta_\bfk$ by $\pi$. Calculations like those presented in Fig. S6 lead to the wavevectors of the efficient pump and signal pairs. Here $B_\uz=23\text{ mT}$.             
	}
	\label{figs5}
\end{figure}
\newpage
\begin{figure}[h!]\centering
	\includegraphics[width=1\textwidth]{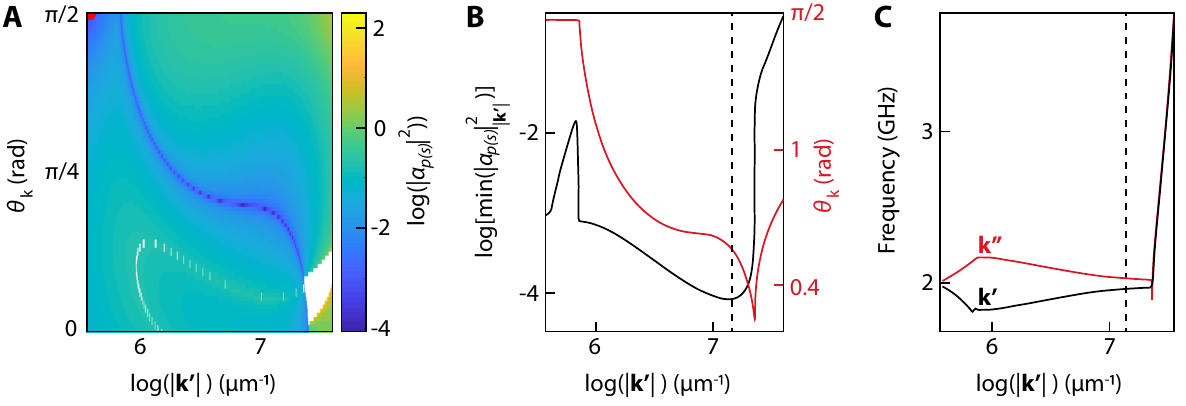}
	\caption{\textbf{Finding the efficient Suhl pair with the lowest excitation threshold.} (\textbf{A}) Calculated threshold amplitude $|\alpha_{\text p (\text s)}|^2$ that triggers the Suhl instability of a spin-wave pair with wavevectors $\bfk'$ and $\bfk''=2\bfk_{\text p (\text s)}-\bfk'$ as a function of $|\bfk'|$ and $\theta_{\bfk'}$. The amplitude is normalized by the total spin $S$. In the white regions no momentum-conserving scattering processes can take place. Here we adopted $B_\uz=24\text{ mT}$, $\omega_{\text p (\text s)}/2\pi=2\text{ GHz}$ and $\theta_{\bfk_{\text p (\text s)}}=\pi/2$, corresponding to the red dot. (\textbf{B}) Minimal threshold amplitude as a function of $|\bfk'|$ (black line, left axis) and corresponding $\theta_{\bfk'}$ (red line, right axis). (\textbf{C}) Frequencies of the modes corresponding to the pairs in (C). The pair with the lowest threshold is indicated by the vertical dashed line in (B) and (C), and defines the “efficient pump (signal) pair” of the Suhl instability.               
	}
	\label{figs6}
\end{figure}
\newpage
\begin{table}[h!]
\centering
\begin{tabular}{ |c|c|c|c|c| } 
 \hline
  Material& $M_\text s$ (A/m) & $D$ (J/m) & $K$ (J/m$^3$) & Reference \\ \hline
 YIG & $1.42\cdot10^5$ & $4.15\cdot10^{-12}$ & $0$ & Ref.~\cite{Bertelli2020} \\ \hline
Permalloy (Py) & $8.46\cdot10^5$ & $2.4\cdot10^{-12}$ & $0$ & \\ \hline 
Cobalt (Co) & $13\cdot10^5$ & $2.4\cdot10^{-12}$ & $0$ & \\ \hline 
CrBr$_3$ & $2.55\cdot10^5$ & $1.2\cdot10^{-12}$ & $2.24\cdot10^5$ & Ref.~\cite{Shen2021} \\ \hline 
CrI$_3$ & $2.15\cdot10^5$ & $1.35\cdot10^{-12}$ & $6.30\cdot10^5$ & Ref.~\cite{Shen2021} \\ \hline 
 Fe$_3$GeTe$_2$ & $3.76\cdot10^5$  & $9.5\cdot10^{-13}$ & $1.46\cdot10^6$ & Ref.~\cite{Leon-Brito2016} \\ 
 \hline
\end{tabular}
\caption{\textbf{Values of the saturation magnetization ($\bm{M_\text{s}}$), exchange constant ($\bm{D}$) and uniaxial anisotropy constant ($\bm{K}$) used to calculate the spin-wave dispersions in Fig. 4F of the main text.}}
\end{table}
\newpage
\printbibliography
\end{refsection}
\end{document}